\documentclass[11pt, letterpaper]{article}
\usepackage[utf8]{inputenc}
\usepackage[margin=1.5cm]{geometry}
\usepackage{titlesec}
\usepackage{tabu}
\usepackage{enumitem}
\usepackage{amssymb}
\usepackage{xcolor}
\newlist{selectlist}{itemize}{2}
\setlist[selectlist]{label=$\square$,leftmargin=*,noitemsep,topsep=0pt}

\usepackage{lmodern}

\usepackage{hyperref}
\hypersetup{
    colorlinks=true,
    linkcolor=blue,
    filecolor=magenta,      
    urlcolor=blue,
}
 
\usepackage{graphicx, import}

\usepackage{biblatex}
\usepackage{array}
\usepackage{subcaption}
\usepackage{makecell}
\newcolumntype{P}[1]{>{\centering\arraybackslash}p{#1}}

\titleformat{\section}[block]{\hspace{1em}\bfseries}{\thesection.}{0.5em}{} 
\titleformat{\subsection}[block]{\hspace{1em}}{\thesubsection}{0.5em}{}
\begin{document}

\begin{flushleft}

{\Huge \bf {\it HardwareX} article template} \textbf{Version 2 (June 2021)}\\
\vskip 0.5cm

\setlength{\parindent}{0pt}
\setlength{\parskip}{10pt}

\textbf{Article title}\\ A Hardened CO2 Sensor for In-Ground Continuous Measurement in a Perennial Grass System 

\textbf{Authors}\\ 
Bobby Schulz, Bryan Runck, Andrew Hollman, Ann Piotrowski, Eric Watkins



\textbf{Abstract}\\ 
Carbon dioxide levels below the soil surface are an important measurement relating to plant health, especially for plants such as perennial grasses in northern climates where ice encasement can occur over winter. In such cases, the CO2 levels can build up and become toxic. This is likely a significant contributor to turfgrass death over winter; however, there is an insufficient amount of data regarding this phenomenon in large part due to the lack of effective sensors. Many off the shelf CO2 sensors exist, but they are not sufficiently hardened for in ground deployment over winter. As a result, the only options currently available are very costly automated gas samplers or manual sampling at intervals with laboratory testing -- a process that results in a limited number of data points and is labor intensive. To combat this problem we have taken an established NDIR CO2 sensor and hardened it for use in winter and ice encased environments to allow for continuous automated sampling of subsurface CO2 levels to better understand ice encasement damage in perennial grass systems. 


\newpage
\textbf{Specifications table}\\
\vskip 0.2cm
\tabulinesep=1ex
\begin{tabu} to \linewidth {|X[1,l]|X[3,l]|}
\hline  \textbf{Hardware name} & \textbf{Hedorah CO2 Sensor}
  \\
  \hline \textbf{Subject area} & %
  \vskip 0.1cm
  \begin{itemize}[noitemsep, topsep=0pt]
  \item \textit{Environmental, planetary and agricultural sciences}
  \end{itemize}
  \\
  \hline \textbf{Hardware type} &
  \vskip 0.1cm  
  \begin{itemize}[noitemsep, topsep=0pt]
  \item \textit{Field measurements and sensors}
  \end{itemize}
  \\ 
\hline \textbf{Closest commercial analog} &
  \href{https://www.seeedstudio.com/SOLO-CO2-5000-A2-p-4857.html}{Seeed Studio SOLO CO2}
  \\
\hline \textbf{Open source license} &
  CC-BY-SA 4.0
  \\
\hline \textbf{Cost of hardware} &
  $<$\$100 USD
  \\
\hline \textbf{Source file repository} & 
      \href{https://zenodo.org/records/13381321}{https://zenodo.org/records/13381321}
\\\hline
\end{tabu}
\end{flushleft}
\newpage
\section{Hardware in context}

Perennial grasses play an important role in sustainable agriculture being used as a cover crop to reduce soil water loss as well as erosion and, in extreme cases, desertification \cite{fan_mechanisms_2020}. Several grasses also are used in managed turfgrass systems including lawns, parks, roadsides, and sports surfaces.
Climate change is likely to result in more volatile environments which put more strain on turfgrasses than in the past. One of the conditions which turfgrass must increasingly contend with is that of winterkill - the inability of the plant to survive the winter, especially in northern climates. The causes of this condition are not completely understood due to the difficulty of monitoring the environment of the turfgrass under ice and snow cover. Winterkill is believed to be caused by a combination of factors including, but not limited to: anoxia, temperature changes when the plant is ill prepared, and CO2 toxicity \cite{valverde_field_2007}. 
While the temperature effects may be difficult to understand as they relate to the plant physiology, they are easily monitored using conventional and well established methods. 

Gas sensing, however, is not as straightforward. The key gases at work in this situation are CO2 and O2. O2 fuel cell based sensors are commercially available and exist in configurations which are suitable for subsurface deployment (e.g. Apogee SO series). CO2 sensors, on the other hand, are not as readily available for the desired use case. 

The most general way to measure CO2 is with some form of spectroscopy. This method relies on the fact that different gasses have different absorption spectra, and if observing the absorption of a single wavelength or a set of wavelengths information can be gained about the concentration of a given gas. For CO2 sensing, the most common version of this is known as NonDispirsive InfraRed (NDIR). In this case a narrow spectral band of IR light is shown through the sample of gas, and the resultant absorption correlates with the CO2 concentration. 

While the principal is simple, the execution has many complications. As a result, high performance NDIR based sensors are still relatively large and need to be placed at the soil surface (e.g. LiCor LI-830) and run sampling tubes down below the soil surface. This limitation means the addition of pumps and other moving parts to bring the gas to the sensor. This complexity leads to an expensive system - routinely costing in excess of \$5000 USD \cite{carbone_flux_2019}. 

One way to combat these problems is to perform manual gas sampling with a syringe or other capture device, then take these samples and perform laboratory testing to evaluate their CO2 concentration, using methods such as liquid gas chromatography \cite{mondini_simple_2010}.  
This method reduces the cost of instrumentation significantly and also eliminates the restriction of the number of sampling sites since the in-ground infrastructure is low in cost. However, this approach requires a technician to travel to each site and collect samples at regular intervals. This need for personnel on site significantly limits the frequency at which samples can be taken.

Using either of these conventional methods the number of sampling sites and the frequency of collection which can be achieved is limited. For a problem like winterkill, where the effects are highly spatially variable and unpredictable, it can make it very difficult to effectively monitor conditions which lead to such a result \cite{valverde_field_2007}.

To resolve this issue we aimed to develop a hardened version of a compact, low cost, NDIR CO2 sensor which could be directly placed in the ground. Our hope was to reduce the cost of the system and therefore increase the available data. 

While this device was designed for use in turfgrass monitoring, there are many areas in which a hardened CO2 sensor can be a useful monitoring tool such as grain storage, produce cultivation/transport, urban/rural air quality, and carbon storage \cite{wang_downhole_2016,neethirajan_carbon_2009,damen_health_2006,maier_monitoring_2010}.

\section{Hardware description}



The design which is presented here is the culmination of several design iteration cycles during which devices were deployed to the field for entire seasons, then the results and the performance were evaluated and modifications made to the design. The final result is the 3rd total iteration of the design - known as the Hedorah-NDIR v2. 

\begin{figure}[!h]
\footnotesize
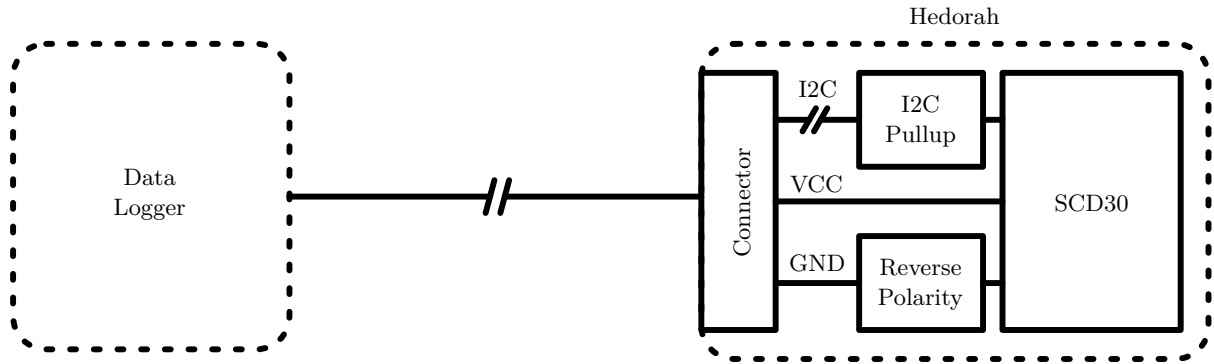
\caption{Block diagram of Hedorah sensor}
\label{fig:BlockDiagram}
\end{figure}

\subsection{Sensor}
Initially an Equivalent CO2 (eCO2) sensor was used (Sensirion SGP30) due to the desirable features of lower cost, power consumption and smaller form factor compared to an NDIR sensor. eCO2 sensors directly measure hydrogen and ethanol concentrations and estimate a CO2 concentration based on these associated gasses. While this method works well for CO2 sources such as human and animal respiration \cite{neethirajan_carbon_2009}, it was found in field testing that this type of sensor was unable to detect the CO2 sources present in soil. Because of this, in the subsequent iteration the design was switched to an NDIR based sensor -- which directly measures CO2 concentration. 

The Sensirion SCD30 was selected for this purpose. This specific device was chosen due to the relative low cost, wide operating range, compact form factor and an electrical contact scheme convenient for manufacturing. The relevance of the device form factor is discussed in greater detail in Sec. \ref{sub:enclosure}. 

\subsection{Connector}
The choice of the way in which the sensor is connected to the data logger is an important consideration. Over many field seasons of deployments with various sensors we observed a frequent issue with damage from coyotes and various rodents. While the sensors themselves were buried, the exposed cables would be damaged significantly in these events. Several methods are used in the community to discourage rodent damage - hot pepper deterrents, mesh enclosures, etc. However, no method was found to be completely effective \cite{shumake_repellents_1999} and the cost of this type of damage was seen to be significant. In our cases the sensor cost was sufficient that they could not be treated as disposable, so the damage needed to be repaired. While the damage was repairable by splicing of the cable without significant materials cost, the personnel cost was significant as a skilled individual was required to make these repairs.  

To get around these issues we desired to replaced the fixed cable observed on other sensors (usually glued in at the sensor end) with a cable which would be removable at the sensor and data logger end. Therefore, in the case of damage, the cable would be treated as disposable and could be replaced by less skilled field technicians - reducing both overall cost and downtime. 

The choice of connector was motivated by the availability of off the shelf cables as well as what was standard within the community. The result was an M12, 4 position, circular connector. This highly robust connector type is commonly used in industrial applications, immersion rated, and designed specifically for low voltage communication applications. The connector is configured so that the female sockets are on the cable/logger side - this prevents any accidental shorting of the cable while disconnected from the sensor. 

This bulkhead type of connector has the added benefit of protecting the device from water intrusion even in the case where the cable is damaged or destroyed.

\subsection{Circuitry/PCB}
The SCD30 is attached to a custom made Printed Circuit Board (PCB) which contains the connector for the data logger cable, as well as minimal circuitry. The circuitry on board the PCB is designed simply to protect the device and provide support for the communication protocol. The SCD30 is a packaged sensor which contains much of the circuitry required for the sensors operation. The only additions which are needed are a set of pull-up resistors to support the I2C bus and a MOSFET which is used to protect the sensor if reverse polarity power is applied. The greater purpose of the PCB is to serve as a mounting point for both the connector and the sensor itself. 

The sensor is connected electrically to the PCB by a set of header pins, but it is also mechanically connected to the PCB by a pair of \#0 machine screws. This additional mechanical connection reduces strain on the electrical connections and reduces the likelihood of fatigue failure. 

\subsubsection{Enclosure}
\label{sub:enclosure}
The enclosure on this sensor is the primary feature. It must serve the purpose of protecting the sensor from mechanical damage, water and soil intrusion, while still allowing passage of gas from the outside environment. 

The key to this gas exchange is the use of Gore-Tex (Expanded PolyTetraFluoroEthylene, ePTFE) vents, which allow for movement of gasses through the membrane while blocking liquid. These vents are available in pre-assembled units where they are mounted into a threaded insert for easy use in the final application. Even with the use of these vents, the enclosure still requires particular attention. 

Given the method of equalization is the diffusion of gas across this membrane, the response time of the system is directly correlated with the volume of air inside of the sensor and the surface area of the membrane. 

The goal of this design is both to have a sensor which performs well and can be manufactured at reasonable quantity. As a result, methods of manufacturing which would require significant labor on the part of the end user are disregarded. In the end we settled on the use of 3D printing as the best option. This would allow us to easily tailor the enclosure to the sensor dimensions while also allowing for easy manufacturing at a scale of 10s to 100s - beyond this, an injection moulded solution should be considered.  

While there are many accessible 3D printing options available, we chose to use the less common Selective Laser Sintering (SLS) method of printing. This method produces sufficiently flat surfaces for making gasket seals without additional post-processing and creates a liquid tight print \cite{morales-planas_multi_2018}. The prints are made of robust nylon, which is both weather and impact resistant \cite{kim_waterproof_2023}. While this method is more expensive, it combines the robustness of FDM printing with the liquid tightness and greater precision of SLA printing to perfectly support this use case. SLS printers are unreasonably expensive for a prototyping environment; however, there are several fabrication facilities which produce parts in small quantities for reasonable prices on such machines (e.g. Protolabs, Shapeways, Xometry, etc).

\begin{figure}[!h]
\centering
\begin{subfigure}{.5\textwidth}
\centering
\includegraphics[width=\linewidth]{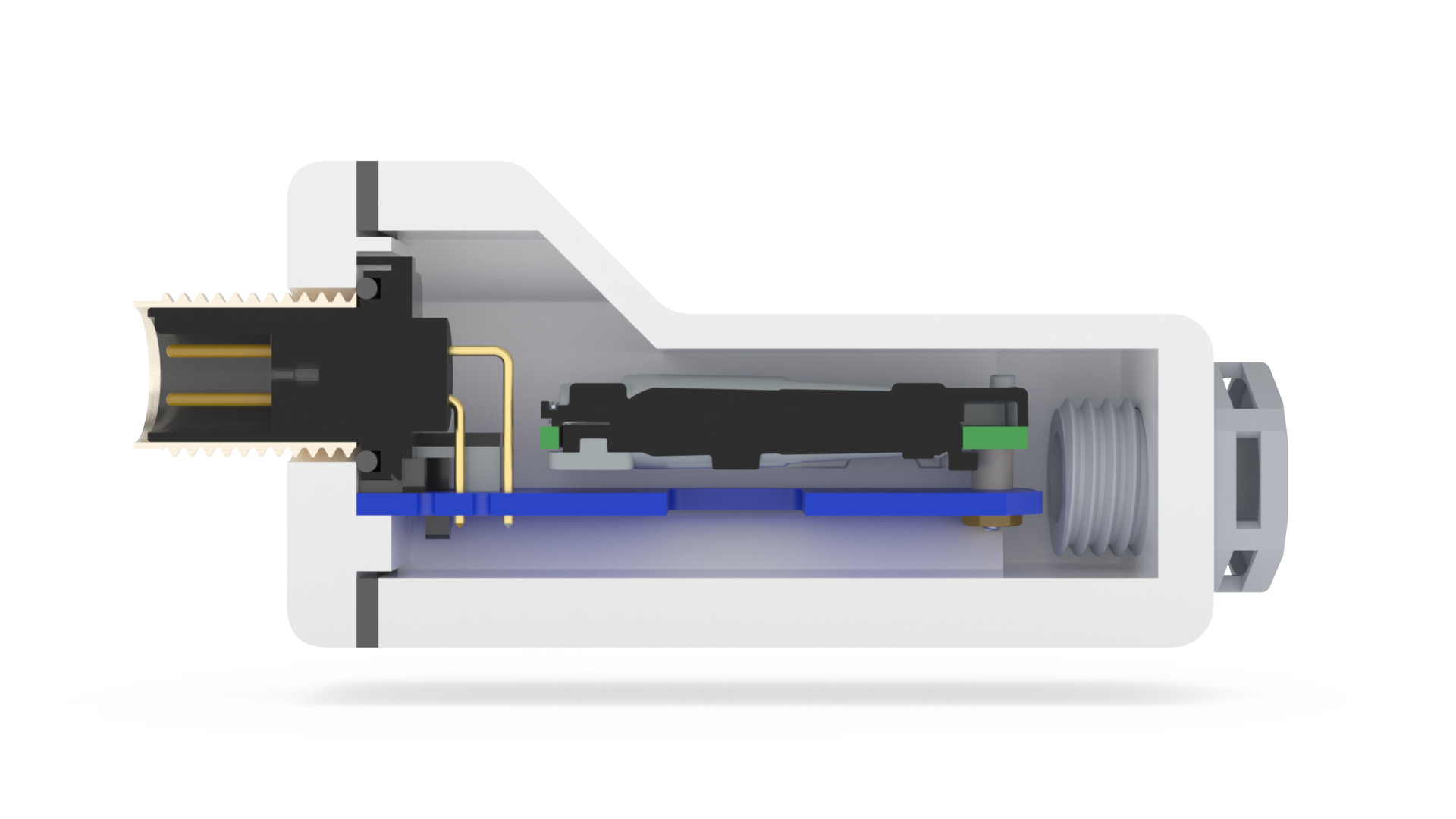}
\caption{Cross section render to showcase internal structure}
\label{fig:HedorahRenderCross}

\end{subfigure}%
\begin{subfigure}{.5\textwidth}
\centering
\includegraphics[width=\linewidth]{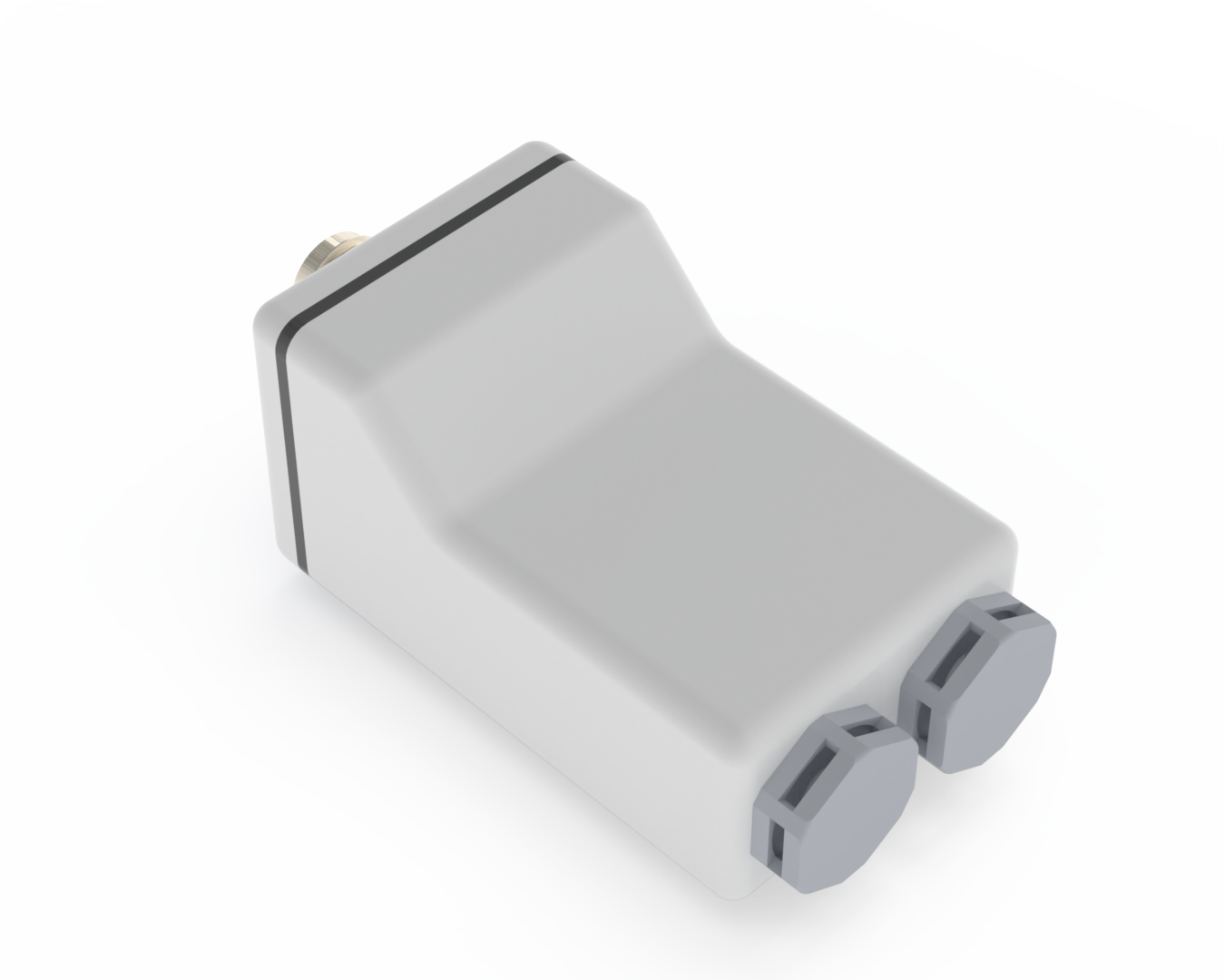}
\caption{External view to show resulted of mated halves}
\label{fig:HedorahRenderISO}
\end{subfigure}
\caption{Renders of various sensor enclosure views}
\label{fig:HedorahRender}
\end{figure}

The final enclosure design is built in two parts, a main body which houses the sensor and mounts the vents, and a lid which seals the main body and allows the connector to pass through. This design is shown in detail in Fig. \ref{fig:HedorahRender}. In this design a face seal gasket is used (shown in detail in Fig. \ref{fig:HedorahRenderCross}) - while this requires the creation of a custom gasket, we found that other methods such as an o-ring face seal or a piston seal were less acceptable. This gasket is cut out of silicone using a laser cutter. Once again, this part can be manufactured by fabrication houses which specialize in small quantity orders (e.g. Ponoko, Xometry, etc). This seal is secured with self tapping screws to reduce need for additional mechanical components such as threaded inserts. 

The vents are secured into the body using threads which are directly printed into the design. 

The key developments of the design are as follows:
\begin{itemize}
    \item Low system cost (device is low cost, and works with low cost data loggers)
    \item Sensor is highly robust - immersion rating unique among CO2 sensors
    \item Easy assembly (can be constructed by nominally skilled workers) and easy to repair (no glue or sealant used in construction)
    \item Wider operating range (0 to 40,000 ppm) than many of the commercially available sensors
\end{itemize}

There are already compact, economical and robust CO2 sensors on the market today (e.g. Seeed Studio SenseCAP SOLO CO2). However, none that we could find offered sufficient robustness for the desired application. In general they are designed to work in dusty or wet environments, at best having an IP rating of 66 - meaning they are rated to be sealed against dust and able to "protect against powerful water jets". This would certainly be sufficient for a sensor which was generally exposed to the elements but in the case of burying these sensors below ground they would be frequently exposed to complete immersion in water which these sensors are specifically not rated to withstand.

\section{Design files summary}
\vskip 0.1cm
\tabulinesep=1ex
\begin{table}
\centering
\begin{tabular}{m{0.25\textwidth} m{0.17\textwidth} m{0.15\textwidth} m{0.25\textwidth}}
\hline
\textbf{Design filename} & \textbf{File type} & \textbf{Open source license} & \textbf{Location of the file} \\\hline
NDIR\_SensorInterface.sch & Eagle CAD schematic file & CC-BY-SA-4.0 & \href{https://zenodo.org/records/13381321}{https://zenodo.org\-/records/13381321} \\\hline
NDIR\_SensorInterface.brd & Eagle CAD board file & CC-BY-SA-4.0 & \href{https://zenodo.org/records/13381321}{https://zenodo.org\-/records/13381321} \\\hline
BetaBody\_v001.SLDPRT & Solidworks \ design file & CC-BY-SA-4.0 & \href{https://zenodo.org/records/13381321}{https://zenodo.org\-/records/13381321} \\\hline
BetaBody\_v001.STL & STL output from Solidworks CAD & CC-BY-SA-4.0 & \href{https://zenodo.org/records/13381321}{https://zenodo.org\-/records/13381321} \\\hline
BetaLid\_v000.SLDPRT & Solidworks design file & CC-BY-SA-4.0 & \href{https://zenodo.org/records/13381321}{https://zenodo.org\-/records/13381321}\\\hline
BetaLid\_v000.STL & STL output from Solidworks CAD & CC-BY-SA-4.0 & \href{https://zenodo.org/records/13381321}{https://zenodo.org\-/records/13381321} \\\hline
BetaGasket\_v001.SLDPRT & Solidworks design file & CC-BY-SA-4.0 & \href{https://zenodo.org/records/13381321}{https://zenodo.org\-/records/13381321} \\\hline
BetaGasket\_v001.DXF & DXF output from Solidworks CAD & CC-BY-SA-4.0 & \href{https://zenodo.org/records/13381321}{https://zenodo.org\-/records/13381321} \\\hline
\end{tabular}
\end{table}

\vskip 0.3cm
\noindent
\begin{itemize}
    \item \textbf{NDIR\_SensorInterface.sch}: Eagle CAD schematic file of the interface board for the sensor. This schematic details the minimal components for the interface board, as well as the pin mapping between the main connector and the sensor itself.
    \item \textbf{NDIR\_SensorInterface.brd}: Eagle CAD board file of the interface board for the sensor. This board contains the minimal set of components required for interfacing and protecting the sensor, while also providing the mounting point for the sensor and the main connector. 
    \item \textbf{BetaBody\_v001.SLDPRT}: Solidworks part file for the main body of the enclosure. This component is the large body shown in Fig. \ref{fig:HedorahRender} and contains the mounting points for the Gore-Tex vents as well as the screws which seal the body at the opposite end.
    \item \textbf{BetaBody\_v001.stl}: STL export of BetaBody\_v001.SLDPRT, generated to be used by 3D printing slicing programs.
    \item \textbf{BetaLid\_v000.SLDPRT}: Solidworks part file for the 'lid' of the enclosure. This component is shown enclosing the end of the 'body' in Fig. \ref{fig:HedorahRender} and allows for passthrough of the main sensor connector as well as the screws to seal the enclosure.
    \item \textbf{BetaLid\_v000.stl}: STL export of BetaLid\_v000.SLDPRT, generated to be used by 3D printing slicing programs.
    \item \textbf{BetaGasket\_v001.SLDPRT}: Solidworks part file for the sealing gasket. This gasket is placed between the 'body' and 'lid', seen as dark band in Fig. \ref{fig:HedorahRender}. This gasket has pass through holes for the screws to clamp the two enclosure parts together and designed to seal the connection between these two pieces.
    \item \textbf{BetaGasket\_v001.DXF}: DXF vector graphic output of the outline of BetaGasket\_v001.SLDPRT, generated to be used by laser cutting processes.
\end{itemize}




\section{Bill of materials summary}
\vskip 0.2cm
\tabulinesep=1ex
\noindent
\begin{tabu} to \linewidth {|X[0.8,1]|X|X[0.6,1]|X[0.8,1]|X|X|X[0.8,1]|}
\hline
\textbf{Designator} & \textbf{Component} & \textbf{Number} & \textbf{Cost per unit - currency} & \textbf{Total cost - currency} & \textbf{Source of materials} & \textbf{Material type} \\\hline

Housing & Body (Body\_v001) & 1 & \$14.8 - USD\footnote{\label{note10}At quantity 10} & \$14.8 - USD & \href{https://www.hubs.com/}{hubs.com} & Polymer\\\hline
Housing & Lid (Lid\_v000) & 1 & \$3.37 - USD\textsuperscript{\ref{note10}} & \$3.37 - USD & \href{https://www.hubs.com/}{hubs.com} & Polymer\\\hline
Housing & Gasket (Gasket\_v001) & 1 & \$4.57 - USD\textsuperscript{\ref{note10}} & \$4.57 - USD & \href{https://www.ponoko.com/}{ponoko.com} & Polymer\\\hline
Sensor & PCB (Hedora-NDIR-0v0) & 1 & \$0.5 - USD\textsuperscript{\ref{note10}} & \$0.5 - USD & \href{https://jlcpcb.com/}{jlcpcb.com} & Composite\\\hline
Housing & Thread Forming Screws, \#6, 3/8", 18-8 Stainless & 4 & \$0.3 - USD\footnote{\label{note50}At quantity 50} & \$1.2 - USD & \href{https://www.mcmaster.com/96001A260/}{96001A260} & Metal\\\hline
Sensor & Machine Screw, 0-80, 5/16", 18-8 Stainless & 2 & \$0.09 - USD\footnote{\label{note100}At quantity 100} & \$0.18 - USD & \href{https://www.mcmaster.com/92196A056/}{92196A056} & Metal\\\hline
Sensor & Spacer, \#0, 1/8", Nylon & 2 & \$0.64 - USD\footnote{\label{note25}At quantity 25} & \$1.28 - USD & \href{https://www.mcmaster.com/94639A459/}{94639A459} & Polymer\\\hline
Sensor & Hex Nut, 0-80, 18-8 Stainless & 2 & \$0.06 - USD\textsuperscript{\ref{note100}} & \$0.12 - USD & \href{https://www.mcmaster.com/91841A115/}{91841A115} & Metal\\\hline
Sensor & Cable, M12, 1.5m\footnote{Length and cable termination may vary by desired application} & 1 & \$12.26 - USD & \$12.26 - USD & \href{https://www.digikey.com/en/products/detail/te-connectivity-amp-connectors/1-2273029-1/5167947}{A120961-ND} & Other\\\hline
Housing & Vent, Gore-Tex & 2 & \$4.22 - USD & \$8.44 - USD & \href{https://www.digikey.com/en/products/detail/amphenol-ltw/VENT-PS1NGY-O8002/7898276?s=N4IgTCBcDaIGoFEByAVAtABQMoEYkHEBNNAeQA4AGCiAXQF8g}{1754-1221-ND} & Polymer\\\hline
Sensor - J1 & Connector, M12 & 1 & \$10.5 - USD & \$10.5 - USD & \href{https://www.digikey.com/en/products/detail/amphenol-conec/43-01204/5144877}{626-2136-ND} & Metal\\\hline
Sensor - J2 & Connector, Male, 4 Position & 1 & \$0.84 - USD & \$0.84 - USD & \href{https://www.digikey.com/en/products/detail/mill-max-manufacturing-corp/800-10-004-10-002000/265875?s=N4IgTCBcDaIBwAYEFoCMKkBY0YWJCIAugL5A}{ED7564-04-ND} & Metal\\\hline
Sensor - Q1 & MOSFET & 1 & \$0.43 - USD & \$0.43 - USD & \href{https://www.digikey.com/en/products/detail/diodes-incorporated/dmg3404l-7/5223202}{DMG3404L-7DICT-ND} & Semi\-conductor\\\hline
Sensor - R1/R2 & Resistor, 10k$\Omega$ & 2 & \$0.1 - USD & \$0.2 - USD & \href{https://www.digikey.com/en/products/detail/yageo/RC0603FR-0710KL/726880}{311-10.0KHRCT-ND} & Metal\\\hline
Sensor & Sensor, CO2 & 1 & \$36.38 - USD & \$36.38 - USD & \href{https://www.digikey.com/en/products/detail/sensirion-ag/SCD30/8445334}{1649-1098-ND} & Other\\\hline

\end{tabu}\\
\vskip 0.2cm
\noindent
Manufacturing specifications of housing and PCBs are described in detail in \href{https://github.com/GEMS-sensing/DFM_-_Hedorah-NDIR}{DFM\_-\_Hedorah-NDIR}. 
\section{Build instructions}

Before assembly process can proceed, the PCB and all surface mount components must be assembled. During the assembly process, all precautions described in \href{https://sensirion.com/media/documents/6D95AA80/6374D8C1/Sensirion_Handling_Instructions_SHTxx.pdf}{Senserion Handling Instructions For SHTxx Humidity and Temperature Sensors} should be followed. Once the PCB is assembled, the following construction should be performed.

\begin{enumerate}[label={(\alph*)}]
\item Insert \textbf{J1} connector into top (component) side of PCB until board guide clicks in
\item Solder contacts of \textbf{J1} on opposite (solder) side of PCB from connector
\item Insert \textbf{J2} into corresponding holes in top (component) side of previously assembled PCB
\item Insert 0-80 screws (2) through the top (side with pin labels) of the SCD30 module
\item Place spacers (2) on protruding screws
\item Insert screws into top (component) side of PCB, with spacers between SCD30 module and custom PCB - ensure correct orientation by confirming the electrical contacts on the SCD30 module align with the header on the custom PCB
\item Thread 0-80 nuts (2) onto bottom of protruding screws - do not over tighten 
\item Solder \textbf{J2} to both the custom PCB and CO2 sensor 
\item Remove jam nut (if still in place) from \textbf{J1}, set aside
\item Pass barrel of M12 connector (\textbf{J1}) through inside (side with ridge) of \textbf{Lid}
\item Replace jam nut on M12 connector (\textbf{J1}), tighten down to \textbf{1Nm} torque
\item Place \textbf{Gasket} on corresponding lip on inside of lid
\item Slide \textbf{Body} over the previous sub-assembly so that it can mate with the gasket - this will only fit in one orientation
\item Use thread forming screws (4) to secure the Lid to the Body through the holes in the Lid - Tighten to \textbf{0.9Nm} torque
\item Install Vents (2) in threaded holes at end of Body - tighten to \textbf{0.7Nm} torque
\end{enumerate}

\section{Operation instructions}

\subsection{Interface with data logger}
Regardless of the placement of the sensor in the environment the interface to the data logger will remain the same. 

\begin{enumerate}[label={(\alph*)}]
    \item Program data logger with appropriate interface program - See \href{https://github.com/sparkfun/SparkFun_SCD30_Arduino_Library}{Sparkfun SCD30 Arduino library}, \href{https://sensirion.com/media/documents/D7CEEF4A/6165372F/Sensirion_CO2_Sensors_SCD30_Interface_Description.pdf}{Sensirion Interface Description} and \href{https://github.com/Sensirion/embedded-scd/releases/tag/2.1.1}{Sensirion Example Code} for examples on how to interface with the sensor
    \item Wire removable cable to data logger using the pinout specified in Table \ref{tab:pinout}, connecting both power and I2C - make sure not to exceed the voltage limits on these connections\footnote{See \href{https://github.com/bschulz1701/Project-Hedorah}{Github} and \href{https://sensirion.com/media/documents/4EAF6AF8/61652C3C/Sensirion_CO2_Sensors_SCD30_Datasheet.pdf}{datasheet} for details} 
    \item Connect M12 end of cable to sensor
    \item Confirm sensor readings
\end{enumerate}

\begin{table}[]
\centering
\caption{Sensor pinout - assuming standard M12 cable color coding }
\label{tab:pinout}
\begin{tabular}{c|c|c}
\textbf{Pin} & \textbf{Color} & \textbf{Function} \\ \hline
1            & Brown          & SDA               \\
2            & White          & Vin+              \\
3            & Blue           & SCL               \\
4            & Black          & Vin-             
\end{tabular}
\end{table}

\subsection{Sensor placement}
Installation procedures will be provided for the specific use case of turfgrass monitoring, as the details of device placement will vary depending on the environment to be sensed. 

The goal of this particular installation is to place the sensor just below the soil surface (as shown in Fig. \ref{fig:SoilDiagram}), attempting as much as possible to reduce the disruption to the soil environment. 

\noindent
\textbf{Required tools}
\begin{itemize}
    \item Cup Cutter
    \item Metal Pipe, 1.5 inch diameter, $\sim$ 6 inch long
    \item Mallet
    \item '2x4' (1.5 x 3.5 inch) wood board, short ($\sim$ 12 inch) 
\end{itemize}

\begin{figure}[h!]
\centering
\fontsize{18pt}{1em}
\def\svgwidth{\columnwidth}
\resizebox{0.55\textwidth}{!}{
\begingroup%
  \makeatletter%
  \providecommand\color[2][]{%
    \errmessage{(Inkscape) Color is used for the text in Inkscape, but the package 'color.sty' is not loaded}%
    \renewcommand\color[2][]{}%
  }%
  \providecommand\transparent[1]{%
    \errmessage{(Inkscape) Transparency is used (non-zero) for the text in Inkscape, but the package 'transparent.sty' is not loaded}%
    \renewcommand\transparent[1]{}%
  }%
  \providecommand\rotatebox[2]{#2}%
  \newcommand*\fsize{\dimexpr\f@size pt\relax}%
  \newcommand*\lineheight[1]{\fontsize{\fsize}{#1\fsize}\selectfont}%
  \ifx\svgwidth\undefined%
    \setlength{\unitlength}{540bp}%
    \ifx\svgscale\undefined%
      \relax%
    \else%
      \setlength{\unitlength}{\unitlength * \real{\svgscale}}%
    \fi%
  \else%
    \setlength{\unitlength}{\svgwidth}%
  \fi%
  \global\let\svgwidth\undefined%
  \global\let\svgscale\undefined%
  \makeatother%
  \begin{picture}(1,0.66666667)%
    \lineheight{1}%
    \setlength\tabcolsep{0pt}%
    \put(0,0){\includegraphics[width=\unitlength,page=1]{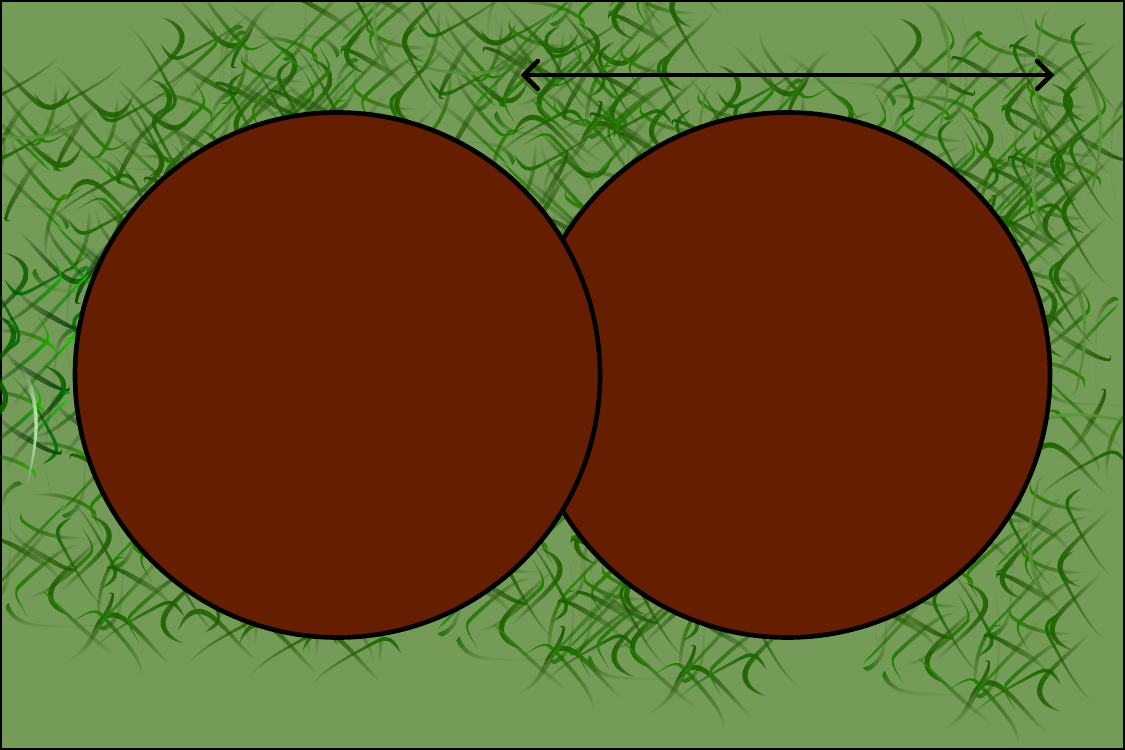}}%
    \put(0.70040693,0.62067057){\color[rgb]{0,0,0}\makebox(0,0)[t]{\lineheight{1.25}\smash{\begin{tabular}[t]{c}4.25 inch\end{tabular}}}}%
    \put(0.5004069,0.03358564){\color[rgb]{0,0,0}\makebox(0,0)[t]{\lineheight{1.25}\smash{\begin{tabular}[t]{c}7.25 inch\end{tabular}}}}%
    \put(0,0){\includegraphics[width=\unitlength,page=2]{CupCut.pdf}}%
    \put(0.08338217,0.60025228){\color[rgb]{0,0,0}\makebox(0,0)[t]{\lineheight{1.25}\smash{\begin{tabular}[t]{c}Turf\end{tabular}}}}%
    \put(0.29999999,0.33333333){\color[rgb]{1,1,1}\makebox(0,0)[t]{\lineheight{1.25}\smash{\begin{tabular}[t]{c}Cutouts\end{tabular}}}}%
    \put(0,0){\includegraphics[width=\unitlength,page=3]{CupCut.pdf}}%
  \end{picture}%
\endgroup%
}
\caption{Diagram cup cut to insert sensor}
\label{fig:CutDiagram}
\end{figure}

\begin{enumerate}[label={(\alph*)}]
    \item Select area for sensor placement 
    \item Use cup cutter to cut and remove a pair of overlapping holes as shown in Fig. \ref{fig:CutDiagram} - cuts should be approximately 6.5 to 7 inches deep
    \item Use the metal pipe to cut and remove a core approximately 3 inches long from the wall at the narrow end of the hole, with the top of the pipe being approximately 2 inches below the soil surface
    \begin{itemize}
        \item Note: If the soil is especially compacted or otherwise difficult to insert the metal pipe, the wooden board can be placed into the hole and the mallet used to strike against it and drive the pipe in
    \end{itemize}
    \item Insert the sensor into the hole created by the metal pipe
    \item Back-fill the hole with the soil and sod from the initial cup cuts, attempting as much as possible to place the soil back in the hole where it originated (this can impact the soil response) - see Fig. \ref{fig:SoilImage} for example result.
\end{enumerate}

\section{Validation and characterization}
\subsection{Example Use - CO2 Monitoring Below Turfgrass}
\begin{figure}[h]
\centering
\begin{subfigure}{.5\textwidth}
\centering
\includegraphics[width=\linewidth,angle=90]{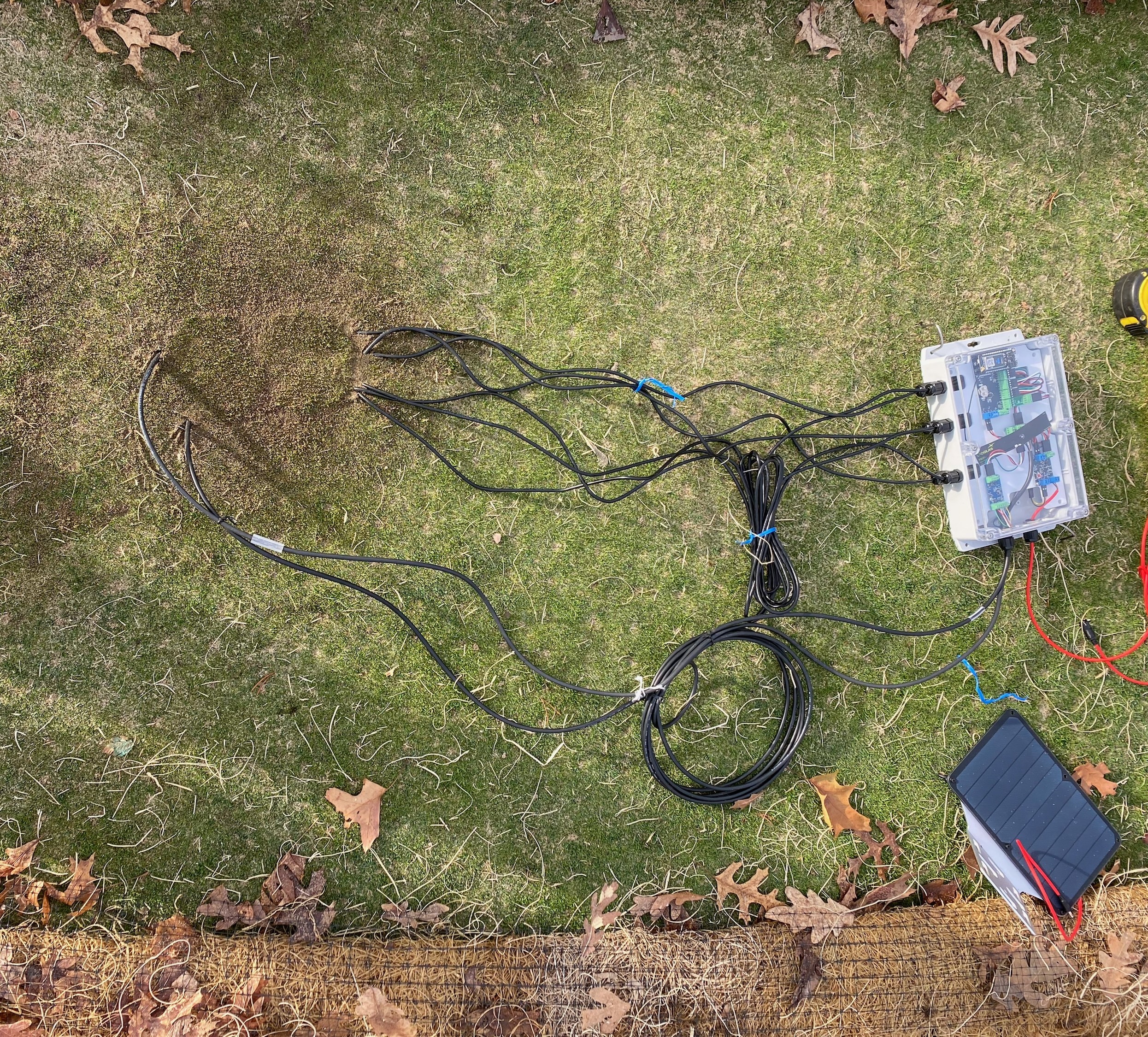}
\caption{Example of sensor placement in field setting. Courtesy of Andrew Hollman}
\label{fig:SoilImage}

\end{subfigure}%
\begin{subfigure}{.5\textwidth}
\centering
\def\svgwidth{\columnwidth}
\resizebox{0.75\textwidth}{!}{
\begingroup%
  \makeatletter%
  \providecommand\color[2][]{%
    \errmessage{(Inkscape) Color is used for the text in Inkscape, but the package 'color.sty' is not loaded}%
    \renewcommand\color[2][]{}%
  }%
  \providecommand\transparent[1]{%
    \errmessage{(Inkscape) Transparency is used (non-zero) for the text in Inkscape, but the package 'transparent.sty' is not loaded}%
    \renewcommand\transparent[1]{}%
  }%
  \providecommand\rotatebox[2]{#2}%
  \newcommand*\fsize{\dimexpr\f@size pt\relax}%
  \newcommand*\lineheight[1]{\fontsize{\fsize}{#1\fsize}\selectfont}%
  \ifx\svgwidth\undefined%
    \setlength{\unitlength}{270bp}%
    \ifx\svgscale\undefined%
      \relax%
    \else%
      \setlength{\unitlength}{\unitlength * \real{\svgscale}}%
    \fi%
  \else%
    \setlength{\unitlength}{\svgwidth}%
  \fi%
  \global\let\svgwidth\undefined%
  \global\let\svgscale\undefined%
  \makeatother%
  \begin{picture}(1,0.88879999)%
    \lineheight{1}%
    \setlength\tabcolsep{0pt}%
    \put(0,0){\includegraphics[width=\unitlength,page=1]{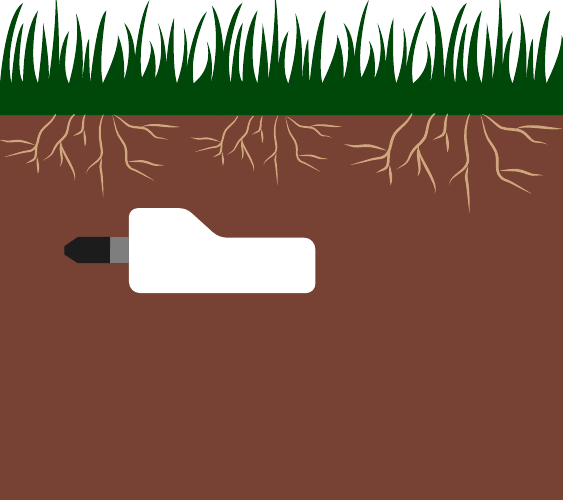}}%
    \put(0.44758786,0.10414828){\color[rgb]{0.73333333,0.73333333,0.73333333}\makebox(0,0)[t]{\lineheight{1.25}\smash{\begin{tabular}[t]{c}\textbf{Soil}\end{tabular}}}}%
    \put(0.39223623,0.39650551){\makebox(0,0)[t]{\lineheight{1.25}\smash{\begin{tabular}[t]{c}\textbf{CO2 Sense}\end{tabular}}}}%
    \put(0.24176147,0.70001675){\color[rgb]{0.73333333,0.73333333,0.73333333}\makebox(0,0)[t]{\lineheight{1.25}\smash{\begin{tabular}[t]{c}\textbf{Turf}\end{tabular}}}}%
    \put(0.75702866,0.5106437){\color[rgb]{0.73333333,0.73333333,0.73333333}\makebox(0,0)[t]{\lineheight{1.25}\smash{\begin{tabular}[t]{c}\textbf{Roots}\end{tabular}}}}%
    \put(0,0){\includegraphics[width=\unitlength,page=2]{SoilSensing.pdf}}%
    \put(0.55433619,0.52800604){\color[rgb]{0.73333333,0.73333333,0.73333333}\makebox(0,0)[t]{\lineheight{1.25}\smash{\begin{tabular}[t]{c}\textbf{2in}\end{tabular}}}}%
    \put(0,0){\includegraphics[width=\unitlength,page=3]{SoilSensing.pdf}}%
  \end{picture}%
\endgroup%
}

\caption{Diagram of sensor under soil surface}
\label{fig:SoilDiagram}

\end{subfigure}
\caption{Example of field installation of CO2 sensor}
\label{fig:CO2Field}
\end{figure}

The primary use of the CO2 sensor has been monitoring CO2 levels below turfgrass, near the root area. This work is aimed at identifying impact of CO2 on winter kill in turfgrass. Fig. \ref{fig:CO2Field} diagrams how the sensor is placed below ground (Fig. \ref{fig:SoilDiagram}) and show the impact above the soil surface (Fig. \ref{fig:SoilImage}). This requires the sensors to operate in a harsh environment with below freezing temperatures, flooded soil, and animal hazards. These sensors have been deployed in these settings for several years across a large fleet and have continued to perform well and instruct design iterations of the sensing device for better performance, serviceability, and manufacturability. 

\begin{figure}[h]
    \centering
    \includegraphics[width=\linewidth]{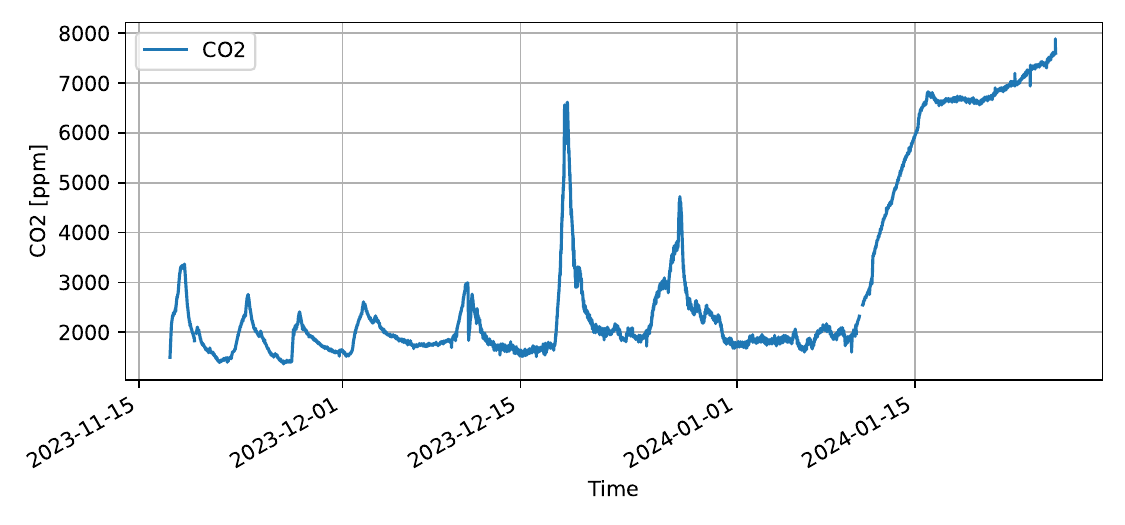}
    \caption{Example below ground CO2 sensing over the course of a winter season}
    \label{fig:FieldData}
\end{figure}

An example of the data collected across a field season is shown in Fig. \ref{fig:FieldData}. From this data we see reasonable fluctuations across the season without any data discontinuities that might indicate a failure of the sensing system. Over the fleet of sensors CO2 values have been observed to range from near atmospheric ($\sim$400 ppm) to saturating the sensor (40000 ppm) and over this range the sensor has continued to perform well. 

Some key features of the design have proven to be especially impactful: the low cost of the sensor has resulted in the ability for a large fleet to be deployed, the small form factor has been beneficial in minimizing the disruption of the soil sub-surface, the detachable cable has been very helpful in the case of animal damage and servicing the device, and the robust enclosure has performed extremely well in the harsh winter environment.

\subsection{Lab Evaluation}
One aspect which is difficult to characterize with field analysis alone is the response time of the system. The steady state performance of the system is specified by the SCD30 manufacturer, however, placing the sensing module into a sealed enclosure which relies on gas exchange for measuring the outside environment will have significant effect on the response time of the system. It is important that this effect is quantified to make sure it is within reason for the application. 

To evaluate this, a chamber must be used which can vary the level of CO2 inside. This was accomplished by following a method described in \href{https://www.mdpi.com/2073-4433/14/2/191#app1-atmosphere-14-00191}{A Low-Cost Calibration Method for Temperature, Relative Humidity, and Carbon Dioxide Sensors Used in Air Quality Monitoring Systems} \cite{gonzalez_rivero_low-cost_2023}. Based on this, a scrubber was first used to reduce the level of CO2 in the chamber below that of atmospheric, then injecting a given volume of CO2 gas rapidly to observe a step in sensor response. From this environmental step response, the time constant\footnote{Time constant, $\tau = 1 - \frac{1}{e} \approx 0.6321$} of the sensor can be found and from this the response time can be closely estimated \cite{liptak_instrument_2003}. 

\begin{figure}[h]
    \centering
    \includegraphics[width=\linewidth]{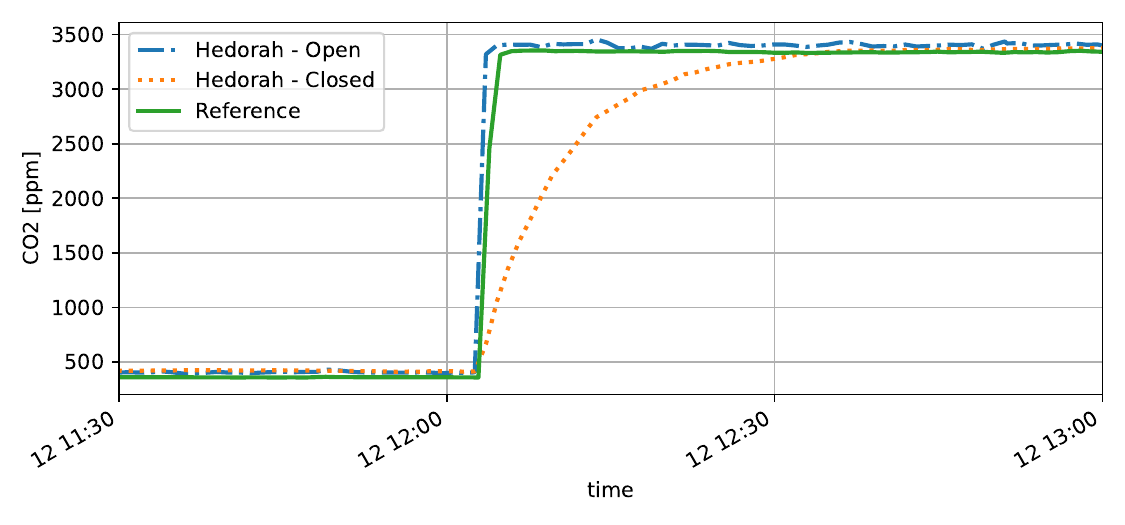}
    \caption{Lab step response evaluation as compared with standard}
    \label{fig:LabData}
\end{figure}

The testing setup consisted of a CO2 sensor with cover removed (open), a CO2 sensor with the cover in place (closed), and lastly a Temtop 2000\footnote{Based on a \href{https://senseair.com/product/s8/}{SenseAir S8}} was included to act as a reference value. The chamber CO2 concentration was stepped from 400 ppm (based on the accuracy range for the reference) to approximately 3500 ppm (the middle of the operational range for the reference). The time response can be observed in Fig. \ref{fig:LabData}. 

It was found that the enclosed sensor had a time constant of approximately 7 minutes, meaning the response time of the sensor (time to reach steady state value) is approximately 35 minutes - well within the expected performance of the system. 

Steady state error was also evaluated by taking a 30 minute mean value both before and after the step response to evaluate the sensor performance for a high and low value. The results are collated in Table \ref{tab:LabData} and in all cases the error of the sensor is within the limits of accuracy of the system.

\begin{table}[h]
\centering
\caption{Results of lab steady state evaluation. Worst case uncertainty is defined as the sum of the accuracy of the test sensor and the reference sensor. }
\label{tab:LabData}
\begin{tabular}{r|ccc}
Condition & \multicolumn{1}{c|}{\textbf{Mean Error {[}ppm{]}}} & \multicolumn{1}{c|}{\textbf{Mean Error {[}\%{]}}} & \textbf{Worst Case Uncertainty {[}ppm{]}} \\ \hline
Open Sensor, before step   & 47 & 12.9 & 93  \\
Closed Sensor, before step & 60 & 16.7 & 93  \\
Open Sensor, after step    & 61 & 1.8  & 272 \\
Closed Sensor, after step  & 38 & 1.1  & 271
\end{tabular}
\end{table}

\clearpage 
\printbibliography

\end{document}